\documentclass[%
 reprint,
 amsmath,amssymb,
 aps,
]{revtex4-2}

\usepackage{graphicx}
\usepackage{dcolumn}
\usepackage{bm}

\begin{document}

\preprint{APS/123-QED}

\title{Conformations and dynamics of active star polymers}

\author{Prabha Chuphal}
 \email{P.Chuphal@skoltech.ru}
\author{Vladimir V. Palyulin}%
 \email{V.Palyulin@skoltech.ru}
\affiliation{%
 Applied AI centre, Skolkovo Institute of Science and Technology, Bolshoy Boulevard 30, Moscow, 121205, Russia.\\
}%




\date{\today}

\begin{abstract}
We study conformations and dynamics of active star polymers. The analysis shows that active star polymers' stretching behaviour is quite different from that of active linear chains. The visual inspection of conformations and bond-bond correlations reveal a better coordination for the alignment and coordination of bonds for the star polymers than for the linear counterparts. The architecture substantially affects the chain extension transition at high values of active force. The scaling laws for the shape factor and the arm asphericity ratio established for the passive star polymers coincide with the passive case for active force values below the transition. For the values above the transition range the scaling of these quantities switches to different values. 
\end{abstract}

\maketitle


\section{\label{sec:level1}Introduction:}

Systems performing motion with energy-consuming elements are known as active matter. Their dynamics essentially happens in non-equilibrium, which effects in a markedly different behaviour compared to their equilibrated counterparts~\cite{RevModPhys:2016}. The study of active matter has opened new avenues for the understanding of various natural phenomena as well as provided means to programme new ways of control and self-assembly ~\cite{Ramaswamy:2010,NatRevPhys:2020}. The active matter systems span a wide range of scales from the microscopic to the macroscopic world, from a self-propulsion in a directed fashion by anchored actin and microtubules over the carpet of motor proteins~\cite{Schaller:2011,Schaller:2010,Prost:2015,Eisenstecken:2016, Winkler:2017,Guimaraes:2020}, an active cytoskeleton of eukaryotic cells~\cite{Brangwynne:2008} to the collective behaviour of groups of animals and bird flocks ~\cite{nuzhin2021,vicsek1995}. Most of the up-to-date research is focussed on agents' behaviour with a small number of degrees of freedom, such as colloidal Janus particles ~\cite{Demirors:2018,Das:2015,Uspal:2019}. Recent studies of more complex systems such as active polymers reveal rich and counter-intuitive dynamics \cite{Sarkar:2016,Anand:2020,Mokhtari:2019,Emanuele:2021,Malgaretti:2018,Chubak:2022,Kurzthaler:2021}. So far the most attention was paid towards active linear polymer chains, both by theory~\cite{Winkler:2017,Malgaretti:2018,tejedor2019reptation,Winkler:2020,Martin:2018} and simulations~\cite{Sarkar:2016,Mokhtari:2019,Anand:2020,Malgaretti:2018,Janke:2022,Mahajan:2022,Jain:2022}. In this case the behaviour is mainly determined by coupling of activity and polymer degrees of freedom~\cite{Winkler:2020}. Similarly, the experimental studies on active polymers also reveal the directed motion of colloidal chain in nematic liquid crystals~\cite{Sasaki:2014}, the aggregation and phase separation of thin living {\it{T. tubifex}} worms through the active motion of worms~\cite{Deblais:2020}. Another layer of complexity comes from passive environment (fluid) itself being able to substantially affect the conformations and dynamics of linear active polymers at low Reynolds numbers~\cite{Kapral:2016,Marchetti:2013,Roland:2019}. In active non-equilibrium media (active fluids) the polymer dynamics  produces a superdiffusive motion at short to intermediate times, which turns normal at long times~\cite{Shin:2017}. The further studies include the discovery of phase separation in active-passive polymer mixtures~\cite{Smrek:2017} with implications for the DNA organisation in living cell nuclei~\cite{Ganai:2014} and diffusion properties of active Ornstein-Uhlenbeck particles in flexible polymer networks~\cite{kim2022active}. In addition, adsorption of an active polymer on a cylindrical surface has also been reported~\cite{Shen:2022}.

Many biopolymers are synthesised or operate in an active environment. For instance, microtubules can be polymerised as well as stabilised by adding some force-generating components, such as kinesin-1 motor proteins. This helps to convert energy from ATP into motion. Mixing of microtubules with motor proteins reduces the spacing between microtubules to a few nanometers to provide the ideal distance for the motors to increase the binding probability~\cite{Nasirimarekani:2021,Marshall:2013}. Self-organisation of microtubules and motor proteins is an exciting feature shown by the biopolymers~\cite{Ndlec:1997,Surrey:2001,Smith:2007}. Some other examples of active biopolymers include cross-linked networks of semiflexible polymers forming a significant part of living tissue and cells~\cite{Broedersz:2014}, active microtubules, or actin filaments responsible for the dynamics of cytoskeleton and chromatin~\cite{Weber:2012,Weber:2015}.

\begin{figure}
\centering
   \includegraphics[width=\columnwidth]{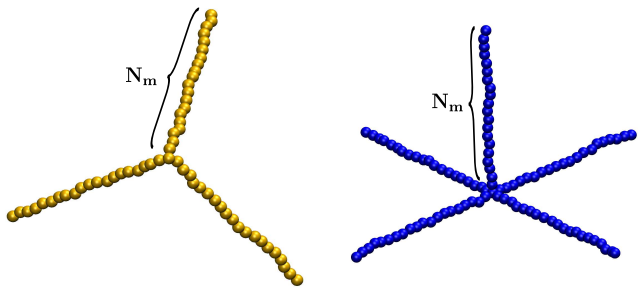}
  \caption{Sketch of active star polymers with three and five arms.}
  \label{intro}
\end{figure}

The topology of conventional polymer molecules is known to span various architectures. For instance, star polymers and other branched polymers (combs, grafts, cross-linked, etc.),~\cite{Yang:2017,comb2009,palyulin2007microphase,comb2007,Maity:2020,potemkin2019,Liffland:2021} are of much interest because in some applications they have advantages over linear polymers due to their tunable parameters. Among these parameters are the number of arms and the length of each arm in the case of star polymers, the number of side chains and side-chain lengths, as well as the backbone length in the case of graft polymers, etc. A star polymer represents a polymer molecule with arms originating from a common centre (Fig.~\ref{intro}). The star polymers are applied for synthesis of nanomaterials~\cite{Ren:2016}, have utilisation in nanotechnology~\cite{England:2020,Yong:2021}, in drug or gene delivery~\cite{Sulistio:2011,Liu:2012,Yang:2017}, nanoreactors~\cite{Gao:2012}, catalysts~\cite{Helms:2005}, could be used in dental applications for the control of bacterial adhesion~\cite{Mortazavian:2019}, bioengineering~\cite{Ren:2016,Monaco:2021,Hayes:2022}, nanoengineering~\cite{Yong:2021}, thermoplastics~\cite{Spencer:2017}, tumor-targeted MRI and chemotherapy~\cite{England:2020,Yang:2021}, and could have an antimicrobial potency~\cite{Salas:2022}. Compared to its linear counterpart, a star-like structure enjoys the advantage of a flexible composition and can carry multiple cargoes within a single molecule. Star-like structures are also used to generate viscoelastic polymer environment for investigating the diffusion of active Brownian particles~\cite{Joo:2020}. 

While the biological systems are inherently active, and many possess branched architectures~\cite{Brangwynne:2008,Weber:2015,Guimaraes:2020}, the synthetic active colloids can also be assembled into chains~\cite{Yan:2016}. In terms of active polymer architectures, apart from linear active polymers, studies of active ring polymers were also performed~\cite{Winkler:2019,Emanuele:2021,Papale:2021,Wang:2021}. A size-dependent collapse and arrest of active ring polymers was investigated by Emanuele {\it{et. al}}~\cite{Emanuele:2021}, where an active self-avoiding ring polymer is constructed by self-propelled monomers. Crazing of non-concatenated ring polymers in the glassy state~\cite{Wang:2021} and ring polymers behaving differently than linear polymers in a mixture of active-passive polymer mixtures~\cite{Papale:2021} have also been reported. The above examples show that the physical properties of polymer structures other than a linear chain could be very different in practice. Despite the practical advantages of  active star polymers, no simulations of active star polymers have been done so far to the best of our knowledge.
In this paper, we explore the effect of activity on such structures and compare the findings with known results of the passive counterparts. A sketch depicting the active stars (with the number of the arms equal to 3 and 5) is shown in Fig.~\ref{intro}. One of our main findings is a noticeable influence of the star architecture on the expansion transition which happens with an increase of the Pe number. In particular, in comparison to the linear polymers the branching point leads to a much better alignment of the bonds and the appearance of coexisting states during the transition.

The article is organised as follows. In section~\ref{model}, we describe our model of the active polymer chain and the simulation parameters. Section~\ref{stars} comprises the overall discussion of results. Section~\ref{conclusion} concludes the findings.

\section{Model}\label{model}
We model active polymers in three dimensions as a string of active Brownian particles (ABP). For excluded volume interactions, Weeks-Chandler-Andersen (WCA) potential for any pair of beads was used,
\begin{equation}
    U_{WCA}(r) = 4\epsilon \left[ \left( \frac{\sigma}{r}\right)^{12} - \left( \frac{\sigma}{r} \right)^{6} \right]+ \epsilon , \hspace{0.5 cm}    r\le r_{c}.
\label{rep_lj}
\end{equation}
Here $\epsilon$ and the diameter of the bead $\sigma$ are the energy and distance parameters, correspondingly. The connectivity of beads into a polymer is ensured by a harmonic potential,
\begin{equation}
    U_{h}=-\frac{k_{s}}{2} \sum_{i=1}^{N-1}\nolimits{(|\textbf{r}_{i+1}-\textbf{r}_{i}|-l_{0})^{2}},
    \label{harmonic}
\end{equation}
where $k_{s}$ is the spring constant of the chain, $l_{0}$ is the equilibrium distance between two beads (bond length), $r_{i}$ and $r_{i+1}$ are the position vectors of the consecutive beads.

The activity is set up by imposing self-propulsion to each bead of a polymer. The direction of the active force is the direction of the velocity $(\hat{\textbf{n}})$ of the individual bead. As in other studies of active polymers the activity is quantified by P\'eclet number~\cite{Emanuele:2021}, $\mathrm{Pe}=\frac{f_{a}\sigma}{k_{B}T}$, where $T$ is a temperature and $f_{a}$ is a magnitude of the active force. Following the Refs.~\cite{Emanuele:2021,Stenhammar:2014} we vary $\mathrm{Pe}$ by changing $k_{B}T$ and fixing $f_{a}=1$ throughout the simulation. This method allows us to approach higher $\mathrm{Pe}$ and avoid possible divergences of force values in simulations. We perform Langevin dynamics simulations using LAMMPS molecular simulation package~\cite{PLIMPTON:1995} for the numerical implementation of the model. The time evolution of the system follows 
\begin{equation}
   \gamma \dfrac{d{\textbf{r}}_{i}}{dt}=  -\nabla U(r) + f_{a}\hat{\textbf{n}}_{i} + \textbf{F}_{r}^{i}.
  \end{equation} 
Here, $\gamma$ is the friction coefficient, $U(r)$ is the effective potential experienced by the polymer monomers, $f_{a}$ is the self-propulsion force applied to $i$-th monomer in the direction $\hat{\textbf{n}}_{i}$ and $\textbf{F}_{r}^{i}$ is a random force describing a thermal motion caused by an implicit solvent at a temperature $T$ and can be obtained from the fluctuation-dissipation theorem as $ \left< \textbf{F}_{r}^{i}(t)\cdot\textbf{F}_{r}^{j}(t^{\prime})\right>=6k_{B}T \gamma \delta_{ij}\delta{(t-t^{\prime})}$.

Orientation of an active monomer $\hat{\textbf{n}}_{i}$ is modelled by the rotational counterpart of Langevin equation,

\begin{equation}
\gamma_{r} \dfrac{d{\hat{\textbf{n}}}_{i}}{dt}=\mathbf{\eta}_{i} \times {\hat{\textbf{n}}_{i}},
\end{equation}
where $\mathbf{\eta}_{i}$ is a torque noise which has zero mean, the variance $\left< \mathbf{\eta}(t)\otimes \mathbf{\eta}(t^{\prime}) \right>=(2k_{B}T)^{2}\delta(t-t^{\prime})/D_{r}$, and $\gamma_{r}=k_{B}T/D_{r}$ is the rotational friction coefficient. $D_{r}$ is the rotational diffusion coefficient and can be expressed in the terms of translational diffusion coefficient $D$ as $D_{r}=3D/l_{0}^{2}$. These overdamped Langevin equations are suitable for modelling the motion in low Reynolds number fluids, for example, in biological environments.

The simulation parameters are taken in dimensionless units. The diameter of a monomer $\sigma=1.0$ sets the unit of length, the mass of a monomer $m=1.0$ sets the unit of mass, and the energy parameter is taken as $\epsilon=1.0$. The equilibrium distance between two adjacent beads is fixed at $l_{0}=1.0$, and the spring constant is chosen to be $k_{s}=10^{3}$. Out of star geometries we mostly study three geometries with arm numbers $n_{a}=3,4,5$. For the stars we checked that for the passive limiting case $(\mathrm{Pe}=0)$ the results agree with the recent theory and simulations of conformational properties of passive stars~\cite{Kalyuzhnyi:2019}. For that we also tested the star with $n_{a}=6,7$ with the number of monomers per arm $N_{m}=10$. However, for other calculations, we chose $N_{m}=50,100$ monomers per arm. Additionally, the model we use reproduces earlier known results for linear active polymers~\cite{Winkler:2017}.  The normalised squared end-to-end distance as a function of $\mathrm{Pe}$ for the number of monomers per chain $N_{m}=100,200$ is plotted it in Appendix A (Fig.~\ref{linear_ply}). One can see that the power-law dependence during the stretching transition follows $R_{e}^{2}\sim \mathrm{Pe}^{2/3}$. This observation is well supported by previous studies~\cite{Winkler:2017,Winkler:2020,Anand:2020}. However, we have studied a wider range of active forces than in the earlier papers and found that for active linear polymers the expansion eventually reaches a plateau (as we will see below the same happens for other architectures (star polymers) as well). We explore the range of active forces from $\mathrm{Pe}=0$ (passive case) to $\mathrm{Pe}=5\times10^{3}$. The time step $\Delta t$ is taken as $10^{-3}$ in all runs. We fixed the density of polymers $\rho\approx10^{-5}$ for all model systems. The dimensions of the simulation box with periodic boundary conditions were set according to the chosen polymer density. Throughout the paper we consider the homogeneous star polymers, i.e. the arms have equal length.

It is important to note that our choice of $\mathrm{Pe}$ number is made according to the choice of Refs. ~\cite{Emanuele:2021,Anand:2020,Anand:2018,Malgaretti:2018} for the sake of more convenient comparison with the results from these papers. It is, however, not a unique choice \textit{per se}. Alternatively, as in Refs.~\cite{Winkler:2017,Eisenstecken:2016, Stenhammar:2014}, one could use the definition $\mathrm{Pe}_R=\frac{f_a}{\gamma D_r\sigma}\sim\frac{1}{k_BT}$. For the case of a single active Brownian particle the characteristic relaxation time is $\tau_R=1/D_r$ and this definition produces $\mathrm{Pe}_R=\frac{f_a \tau_R}{\gamma\sigma}$. For passive flexible linear polymers this relaxation time (Rouse time) reads $\tau_R=\frac{2\gamma N^2\sigma^2}{3\pi k_BT}$~\cite{Doi:1986}. For the passive star polymers the longest relaxation Rouse time is defined by the total length of two longest arms and for the arms of equal length is the same as a linear polymer with $2N_m+1$ monomers~\cite{Doi1974}. However, for active linear polymers the spectrum of relaxation times within the model of Ref. \cite{Eisenstecken:2016} also includes the bond stretching coefficient $\mu$ and scales as $\tau_n=\frac{\tau_R}{\mu n^2}$, where $\mu=4\sigma\lambda/3$, $\lambda$ being activity-dependent bond-stretching coefficient~\cite{Winkler:2017}. In that model $\mu$ itself depends on $\mathrm{Pe}$ for linear polymers and the high activity scaling changes from $\mu\sim \mathrm{Pe}$ to $\mu\sim \mathrm{Pe}^{4/3}$ for short and long chains, correspondingly~\cite{Winkler:2017}. Nevertheless, in our simulation setup we vary the temperature rather than activity itself. The use of definition from Ref. \cite{Eisenstecken:2016} basically means the rescaling of the $\mathrm{Pe}$ value, which will not change the essential observations. In Appendix B we show that our model reproduces the well-known results for passive star polymers while in Appendix C we discuss the gyration radius time correlation function and the difference of relaxation time spectrum in our model versus the model of Ref. \cite{Winkler:2017}.

\section{Simulations and analysis} \label{stars} 
\begin{figure}[h]
\centering
\includegraphics[width=0.95\columnwidth]{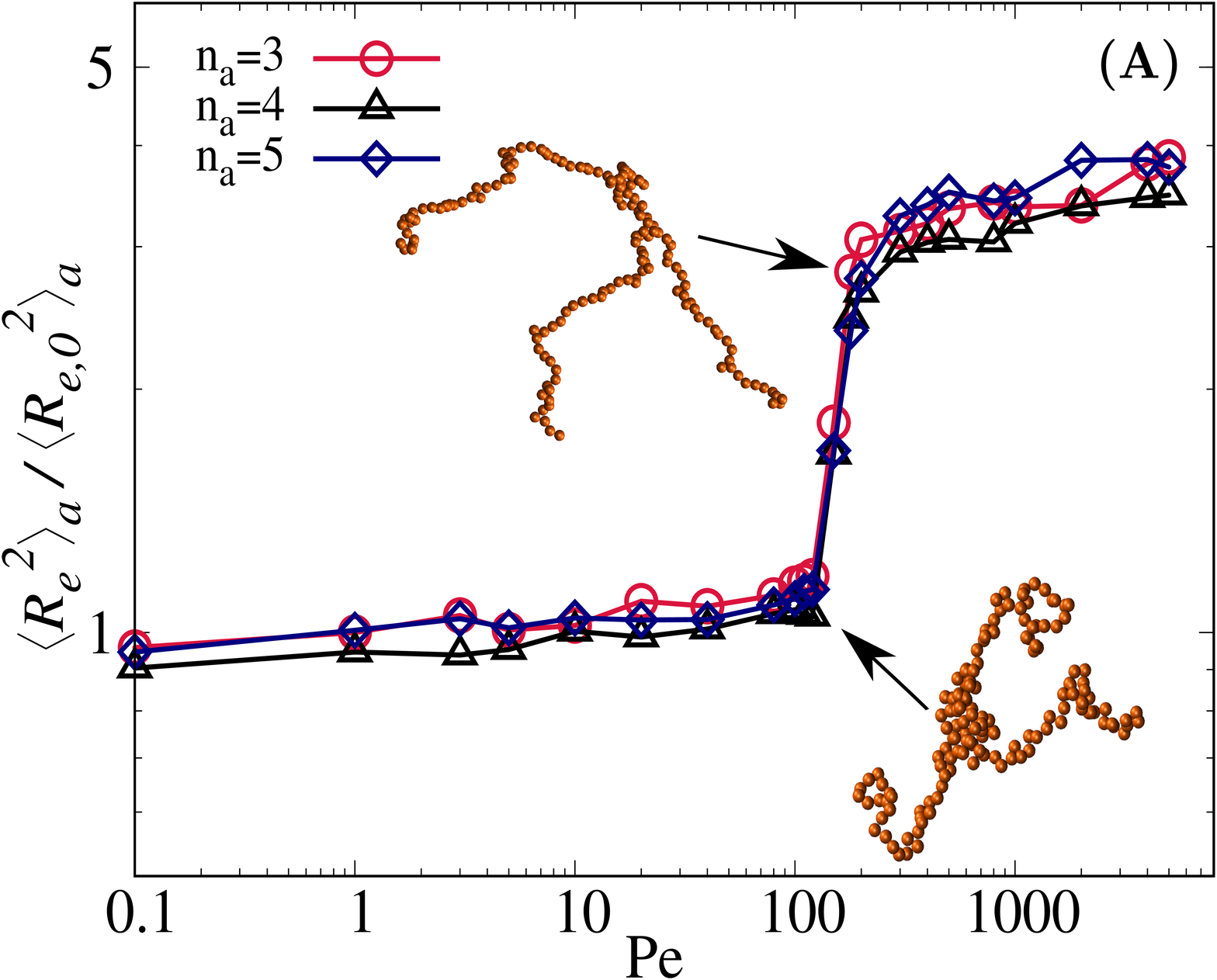}
  \includegraphics[width=0.95\columnwidth]{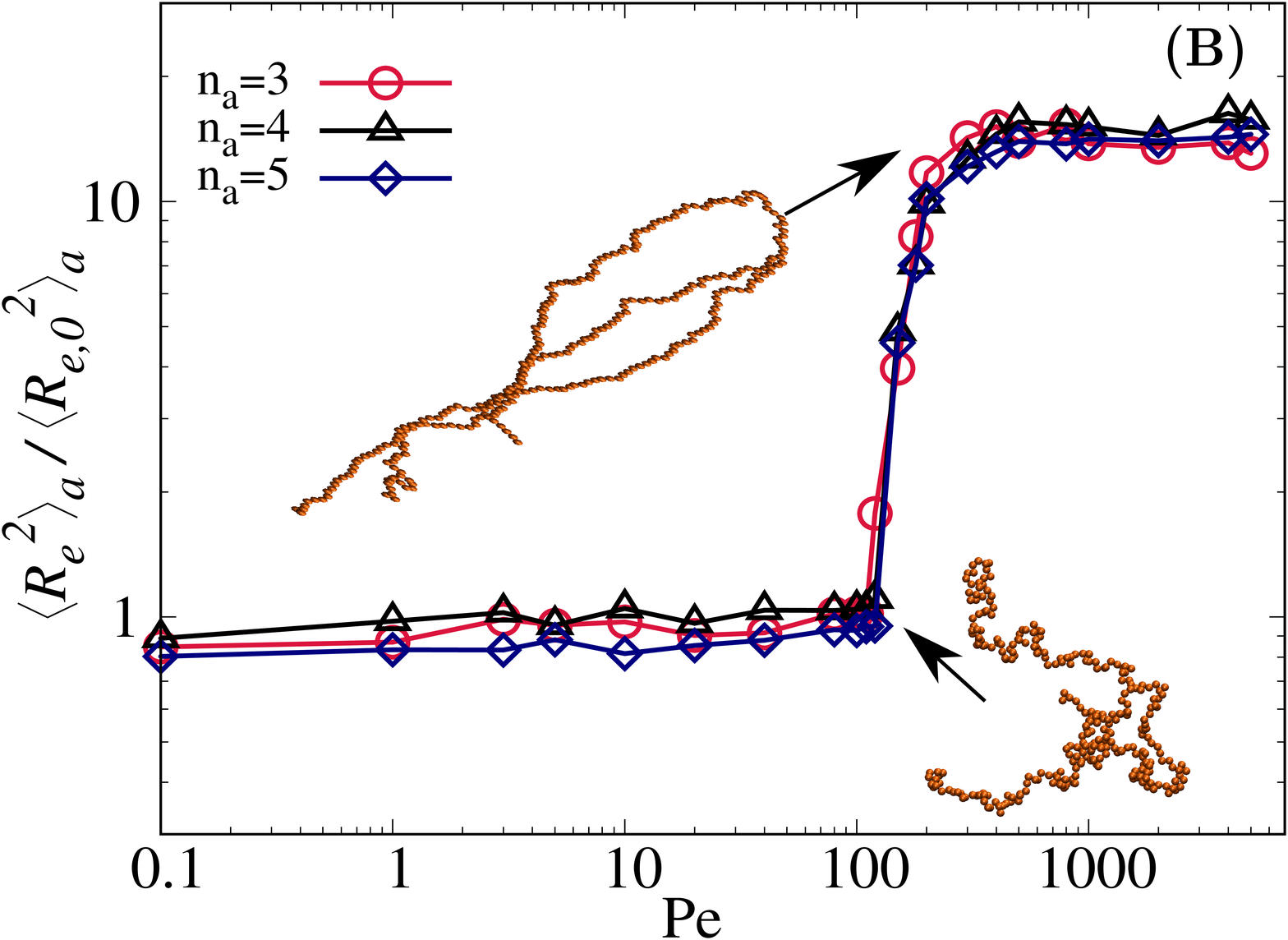}
  \caption{Dimensionless arm stretching ratio $\frac{\left<R_{e}^{2}\right>_a}{\left<R_{e,0}^{2}\right>_a}$ as a function of different magnitudes of active force characterised by $\mathrm{Pe}$ number. Active star polymers with number of arms $ n_{a}=3,4,5$ (circles, triangles, squares, respectively). (A): $N_m=50$. (B): $N_m=100$. In both cases the snapshots of polymer configurations are shown for $\mathrm{Pe}=100$ (more compact) and $\mathrm{Pe}=200$ (more stretched).}
  \label{Re2_stars}
\end{figure}

\begin{figure*}
\includegraphics[width=2.0\columnwidth]{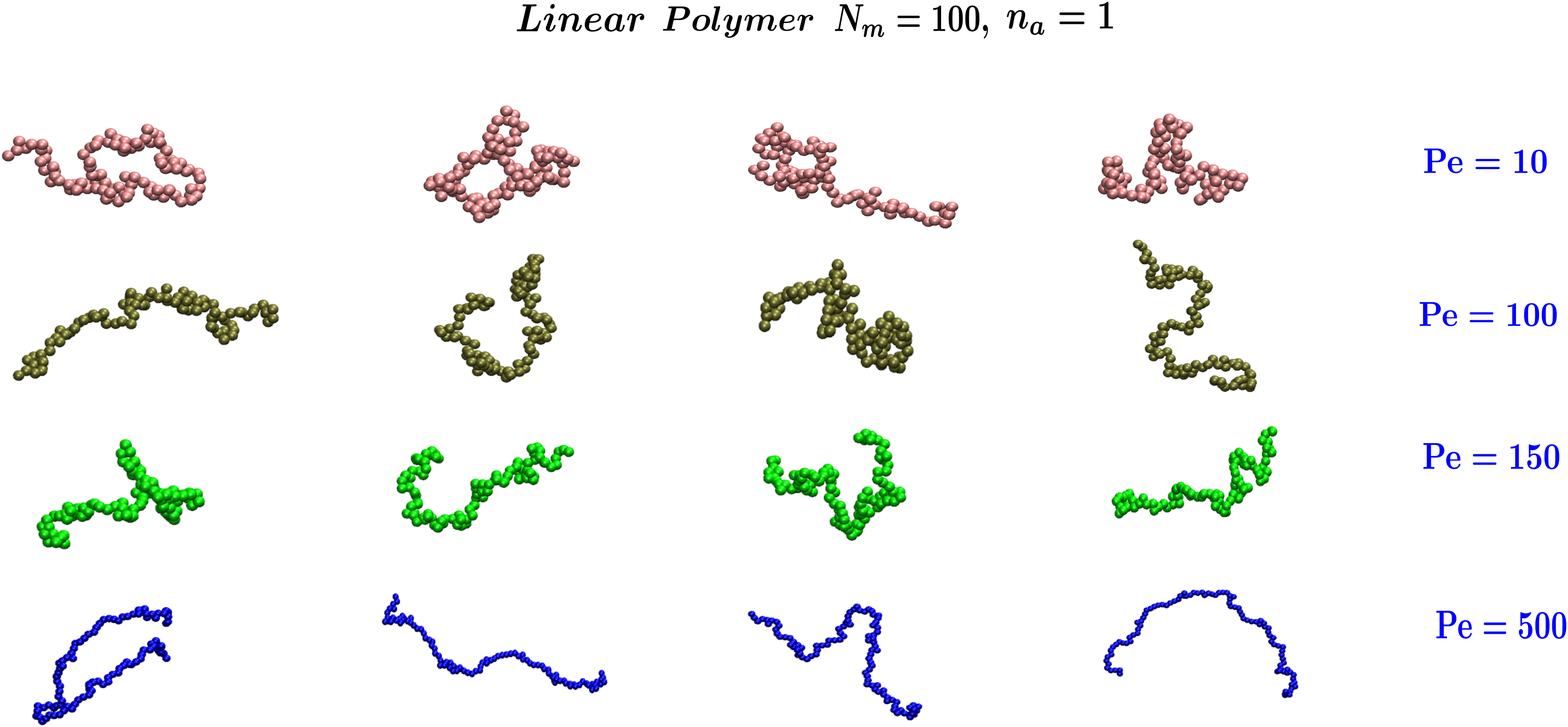}

\vspace{0.5 cm}  \includegraphics[width=2.0\columnwidth]{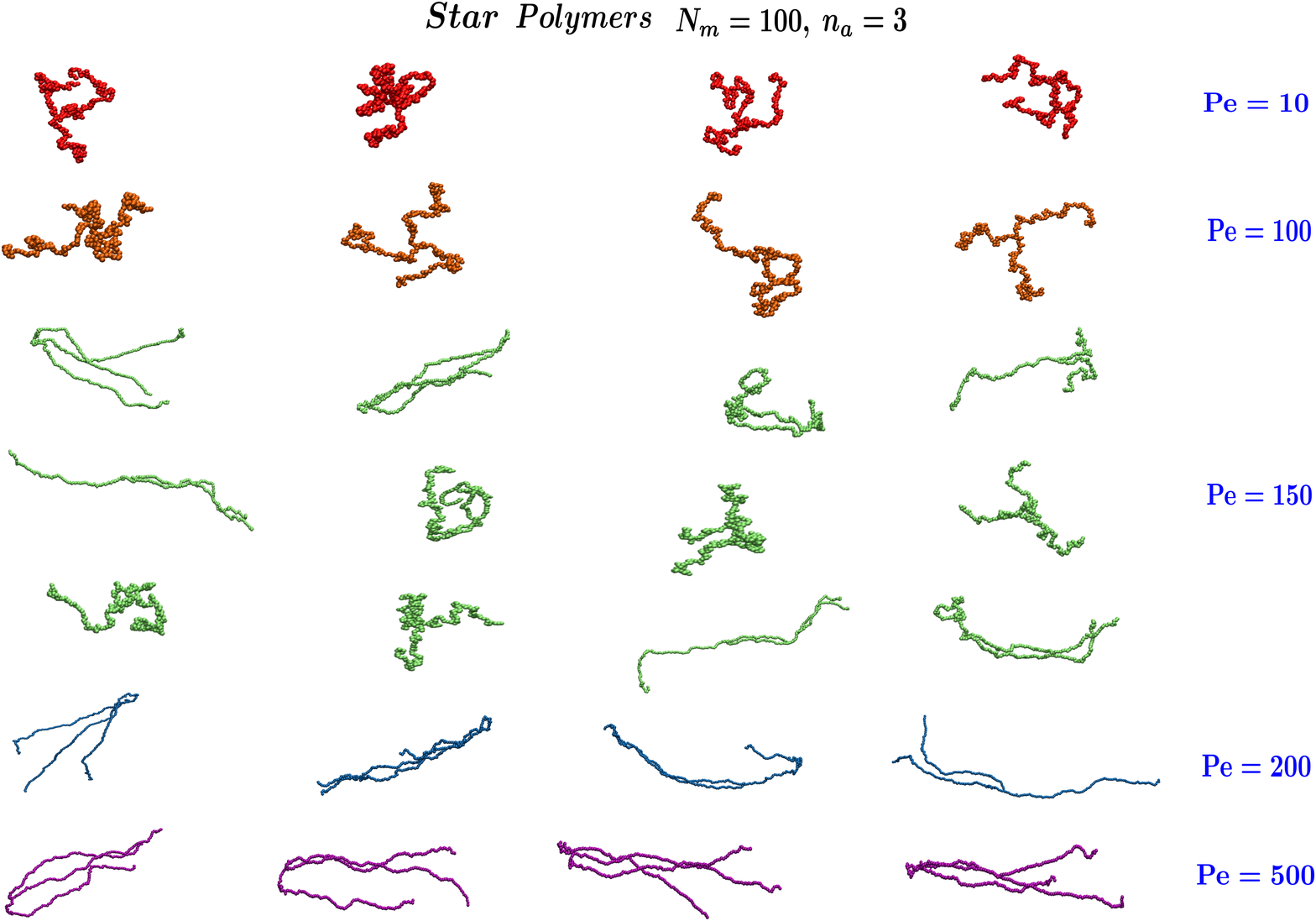}
  \caption{Snapshots showing the typical conformations for activities before, at and after the expansion transition for linear polymer with $N_{m}=100$ and for star polymer with $N_{m}=100$ and $n_{a}=3$. For the case $\mathrm{Pe}=150$ we show more snapshots to illustrate that the star polymers switch from an extended to a collapsed state and back. This explains a probability distribution for the centre-to-end distance $R_e$ of a star polymer shown in Fig.~\ref{prob_star}. Importantly, this coexistence is not observed for the linear polymer (cf. Fig.~\ref{prob_lin}). While in both figures the sizes of polymer chains and the bead diameters are the same, the extended confirmations are pictured in a zoomed out way to be able to fit all snapshots within a single figure.}
  \label{linear_ply_snaps}
\end{figure*}

We first compute the average squared centre-to-end distance of a star polymer, which can be determined as $\left<R_{e}^{2}\right>_a=\left<\frac{1}{a}\sum_{k=1}^{a}[\Vec{\textbf{r}}_{i,N}-\Vec{\textbf{r}}_c]\right>$, where $\Vec{\textbf{r}}_{i,N}$ is the position of $i$-th monomer of $a$-th arm and $\Vec{\textbf{r}}_{c}$ is the position of the central bead. In a specific case of a homogeneous star polymer the average squared centre-to-end distance of a star polymer is equivalent to the averaged end-to-end distance of individual arms.

In Fig.~\ref{Re2_stars} we plot the dimensionless arm stretching ratio $\frac{\left<R_{e}^{2}\right>_a}{\left<R_{e,0}^{2}\right>_a}$ (average squared centre-to-end distance scaled w.r.t. corresponding passive star polymer) as a function of activity (i.e. $\mathrm{Pe}$ values) for two different arm lengths for stars with $(n_{a}=3,4,5)$. Fig.~\ref{Re2_stars}(A) shows the case for $N_{m}=50$ and Fig.~\ref{Re2_stars}(B) is plotted for $N_{m}=100$ monomers per arm. With increase of $\mathrm{Pe}$ the average stretching goes through a transition. Before the transition the stretching is more or less absent. Then a sharp increase in the stretching $\frac{\left<R_{e}^{2}\right>_a}{\left<R_{e,0}^{2}\right>_a}$ is followed by a saturation and a plateau. The expansion is not abrupt and looks linear on the double log scale. However, this linear part spans the range $100<Pe<200-300$ and is too short for the derivation of a scaling dependence. For linear active polymers although the range for the computation of the scaling exponent is rather narrow it produces the values consistent with earlier observations for linear polymers (Appendix A).
It is quite clear though that the slope of the transition region depends on the architecture, in particular on the molecular mass. 





In order to better illustrate the transition differences and similarities to the linear case we perform a visual inspection of the conformation dynamics of star polymers in Fig.~\ref{linear_ply_snaps}. Here we show samples of typical conformations along a single trajectory. One can see that for linear polymers the conformations stretch with an increase of $\mathrm{Pe}$. The coil conformations for low $\mathrm{Pe}$ numbers $(\mathrm{Pe}=10)$ smoothly evolve into extended ones. For star polymers the transition happens through a coexistence of a coil-like and an extended states which is clearly shown for star polymer with three arms at $\mathrm{Pe}=150$ when in time the same polymer changes its conformation from the extended to a coil-like conformation and back. Also it is clear that coordination of monomers for star polymers is considerably stronger than for the linear polymers (cf. case with $\mathrm{Pe}=150$, $\mathrm{Pe}=200$, $\mathrm{Pe}=500$ in Fig.~\ref{linear_ply_snaps}), i.e. an additional geometric constraint improves the overall directional correlations of arms.

\begin{figure}
    \centering
   \includegraphics[width=\columnwidth]{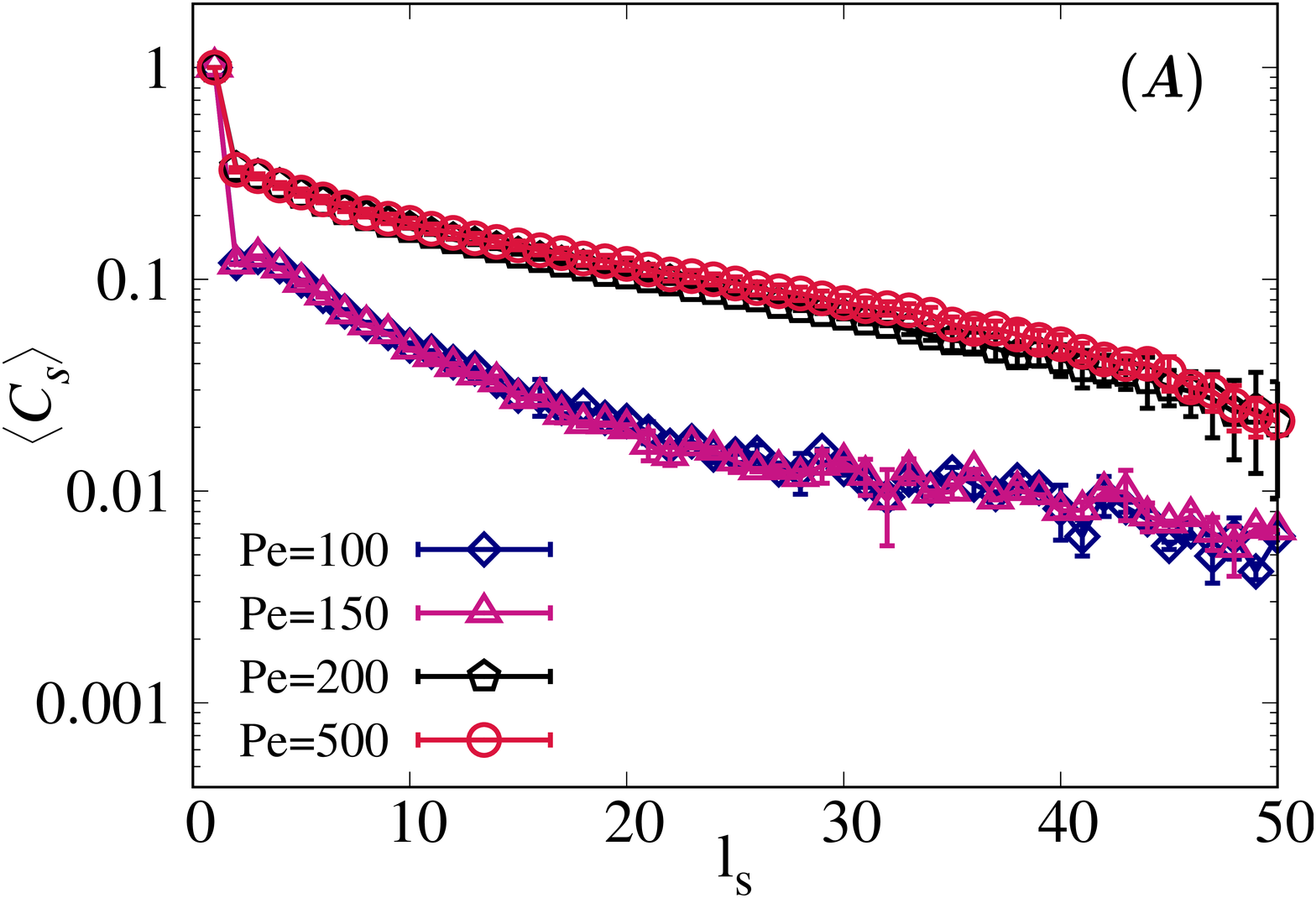}
   \includegraphics[width=\columnwidth]{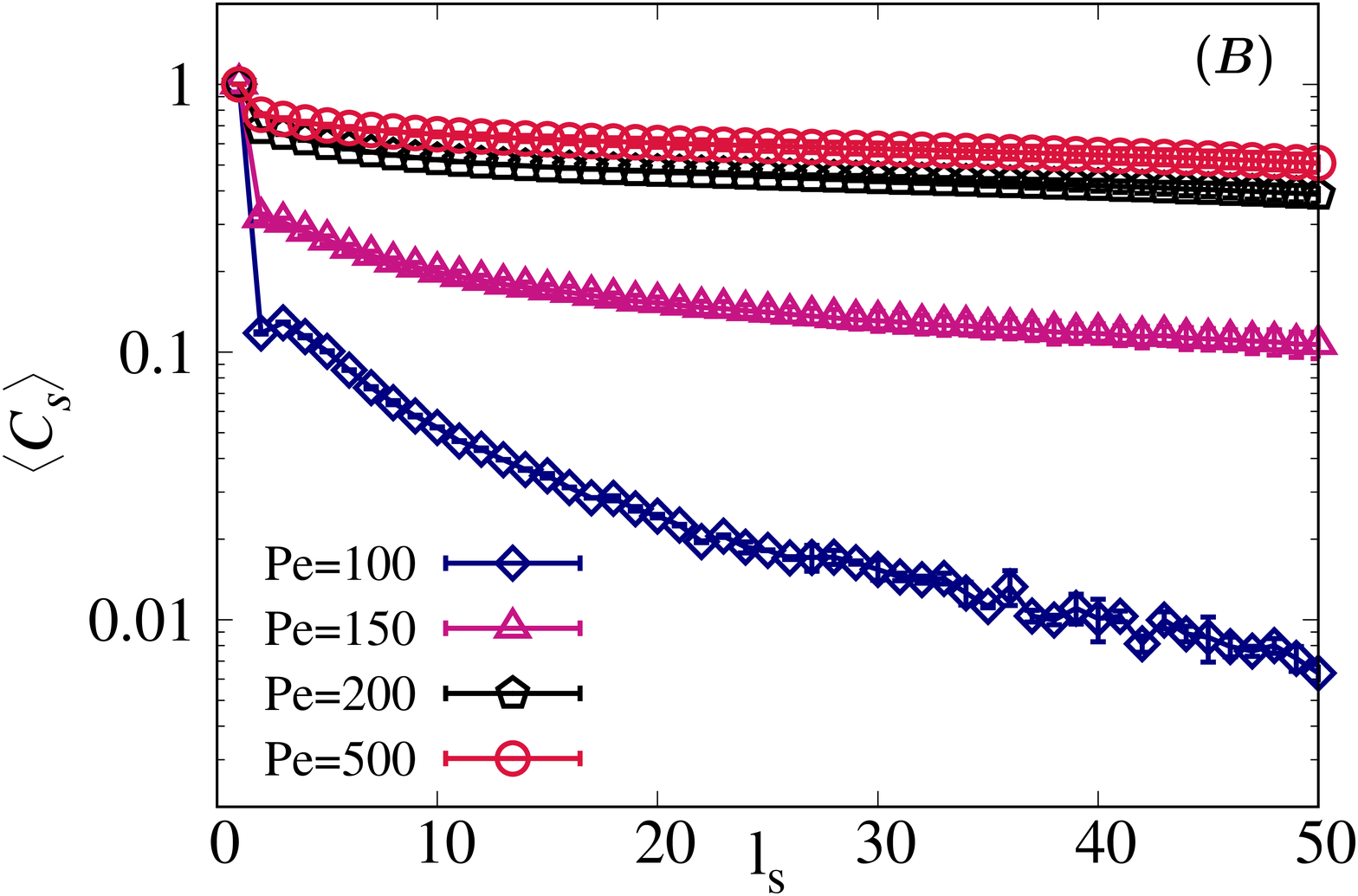}
    \caption{A. Bond-bond correlation for the linear polymer with $N_{m}=100$, B. the bond-bond correlation for the star polymer with with $n_{a}=3$, $N_{m}=100$ on log-linear scale.}
    \label{BB_corr}
\end{figure}

This visual observation is confirmed by the comparative analysis of bond-bond correlations along the chain (see Fig.~\ref{BB_corr}). The correlation of bond directions for active linear polymers (Fig.~\ref{BB_corr}(A)) decays much faster than for active star polymers (Fig.~\ref{BB_corr}(B) for the same $\mathrm{Pe}$ numbers. The bond-bond vector correlations were computed as $\langle C_s\rangle=\langle\mathbf{b}_i\cdot\mathbf{b}_{i+s}\rangle$ from the unit directional vectors of bonds $\mathbf{b}_i$. One can see that the coordination and the alignment of the bonds substantially improved due to joining of linear chains into a star polymer.

\begin{figure}
\centering
 \includegraphics[width=\columnwidth]{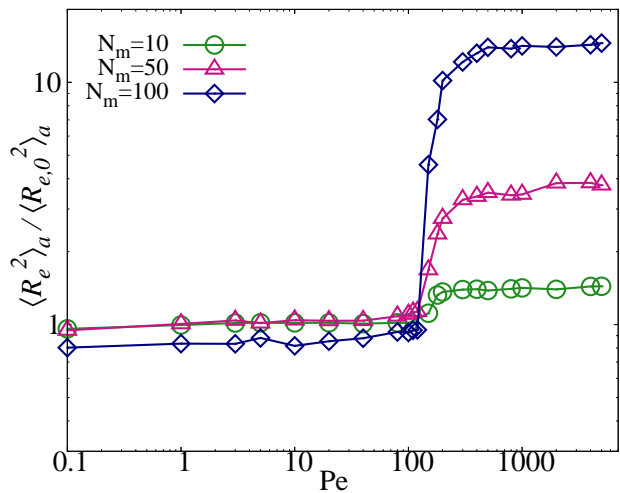}
  \caption{Dimensionless arm stretching ratio $\frac{\left<R_{e}^{2}\right>_a}{\left<R_{e,0}^{2}\right>_a}$ for an active star polymer with the number of arms $n_{a}=5$ and three different number of monomers per arm $(N_{m}=10,50,100)$.}
  \label{Re2_35stars}
\end{figure}

\begin{figure}[h!]
\centering
 \includegraphics[width=\columnwidth]{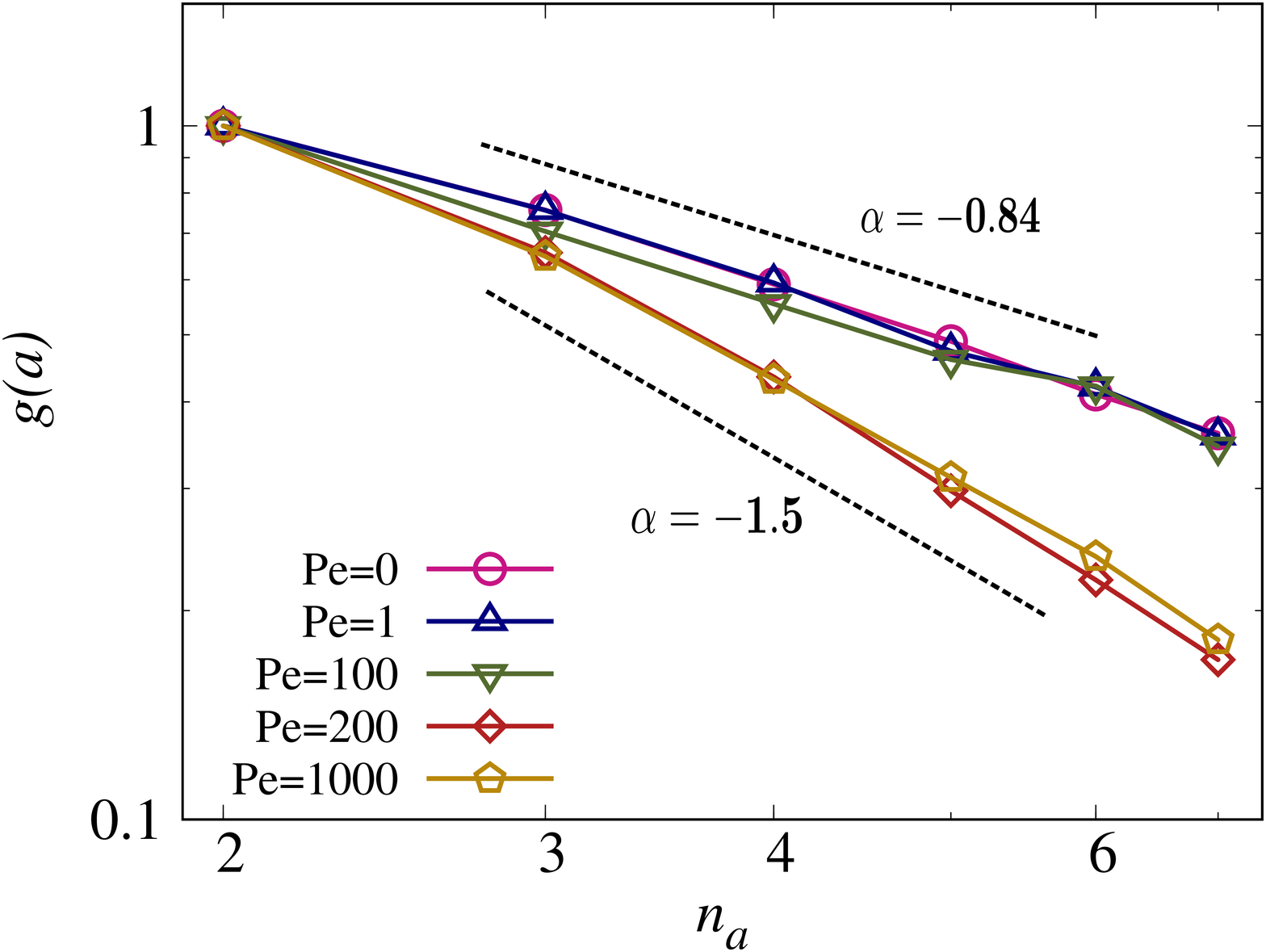}
  \caption{Shape factor $g(a)$ of active star polymers for various activities and different arm numbers $(n_{a}=3,4,5,6)$ and $N_{m}=10$ per arm. The case $n_a=2$ corresponds to a linear polymer. The scaling exponent $0.84$ coincides with the exponent obtained by DPD simulations for passive stars in Fig. 2 of Ref. ~\cite{Kalyuzhnyi:2019}.}
  \label{shape}
\end{figure}

\begin{figure}[h!]
\centering
 \includegraphics[width=\columnwidth]{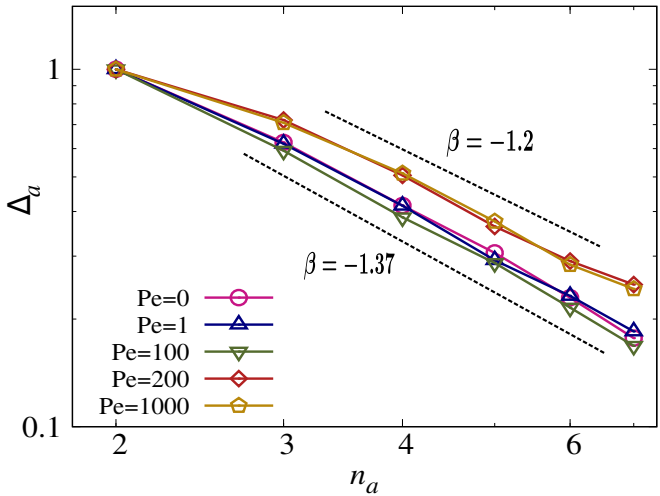}
  \caption{Arm asphericity ratio $\Delta_{a}$ for passive ($\mathrm{Pe=0}$) and active star polymers with monomers $N_{m}=10$ per arm of the star. The scaling exponent $1.37$ is very close to the DPD result in Fig. 3 of Ref. ~\cite{Kalyuzhnyi:2019}.}
  \label{asp_fac}
\end{figure}

\begin{figure}[h]
\centering
  \includegraphics[width=\columnwidth]{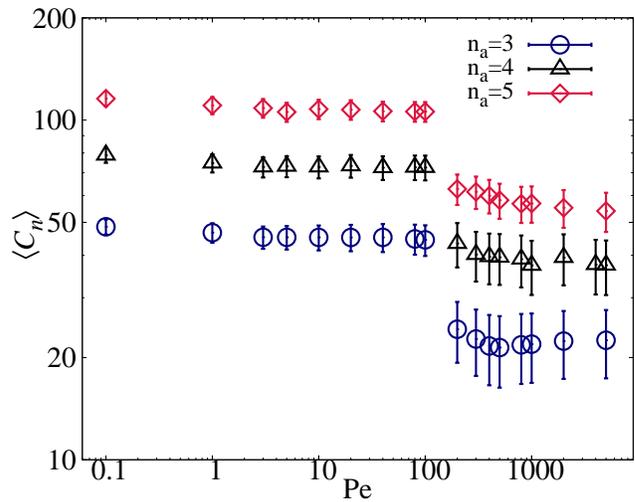}
  \caption{Average number of nearest neighbours of a bead in active star polymers $\left<C_n\right>$ as a function of $\mathrm{Pe}$. Blue circles, black triangles and red diamonds indicate the average count for $n_{a}=3$, $n_{a}=4$ and, $n_{a}=5$, respectively.}
  \label{mono_cnt}
\end{figure}

The effect of an arm's length on the stretching behaviour associated with a particular star polymer is interesting because there are applications based on the molecular weight of stars in the biomedical field~\cite{Wu:2015}. The dimensionless arm stretching ratio $\frac{\left<R_{e}^{2}\right>_a}{\left<R_{e,0}^{2}\right>_a}$ is shown in Fig.~\ref{Re2_35stars} for different molecular weights and $n_{a}=5$. For $n_{a}=3,4$ the plots are qualitatively the same and, thus, are not shown here. In Fig.~\ref{Re2_35stars} we compare three different arm lengths with the number of monomers per arm as $N_{m}=10,50,100$. The shorter arms with monomer number $N_{m}=10$ show smaller expansion when compared to $(N_{m}=50, 100)$. The steepness of the transition from a compact (passive) coil confirmation to the extended one clearly depend on the arm's length rather than the number of arms (cf. Fig.~\ref{Re2_stars}).

The conformation behaviour of passive star polymers is usually described by the shape factor $g_{a}$ and the arm asphericity ratio $\Delta_{a}$. $g(a)$ is defined as $g(a)=\left<R_{g,a}^2\right>/\left<R_{g,1}^2\right>$, where $\left<R_{g,a}^2\right>$ is a squared gyration radius of a star polymer with number of arms $a$  and $\left<R_{g,1}^2\right>$ is squared gyration radius for a linear chain with the same number of monomers. The arm asphericity ratio for a star polymer is defined as $\Delta_A(a)=\left<A_a\right>/\left<A_1\right>$, where $\left<A_{a}\right>$ is the asphericity of the star and $\left<A_{1}\right>$ is the asphericity of the linear polymer with the same molecular mass. For passive star polymers, these quantities have been computed by Kalyuzhnyi {\it{et. al}} with dissipative particle dynamics (DPD) simulations and analytical theory~\cite{Kalyuzhnyi:2019} as well as in some earlier simulations~\cite{whittington1986lattice,Grest:1994,Hsu:2004}. Our simulation results for passive polymers (i.e. $\mathrm{Pe}=0$) are consistent with the data reported by Kalyuzhnyi {\it{et. al}} for the number of arms $n_{a}=3, ..., 7 $ and the number of beads per arm $N_{m}=10$ (Figs.~\ref{shape},~\ref{asp_fac}). The shape factor $g_a$ for active polymers stays the same as for passive counterparts for $\mathrm{Pe}$ values lower than a stretching transition $(\mathrm{Pe=1, 100})$  and markedly changes for higher values $(\mathrm{Pe=200, 1000})$, Fig.~\ref{shape}. For the arm asphericity ratio $\Delta_{a}$ the behaviour also changes due to stretching transition (Fig.~\ref{asp_fac}). For low $\mathrm{Pe}$ values the scaling exponent as a function of number of arms corresponds to the passive case while for high values it decays slower with $a$ although with a close exponent for 3-7 arm range. Thus, we see a good match of scaling laws from Kalyuzhnyi {\it{et. al}} for the passive stars at low $\mathrm{Pe}$ numbers and a quantitative difference at higher $\mathrm{Pe}$ values.

In addition, we check the stretching patterns of different stars with the same overall number of beads (molecular mass). To that end we simulate two cases with two molecular masses, $\approx 51m$ and $\approx 201m$, respectively. The stretching behaviour is shown in Fig.~\ref{Re2_51_201} in Appendix B. A similar trend of initial increment and later saturation in the stretching pattern is observed. One can see that the stretching behaviour is defined by the length of the arm rather than the overall molecular mass of a polymer. That is a consequence of a fact that a star polymer is a set of linear polymers attached together to a common centre. The individual arms stretch in a more or less the same direction and, thus, the stretching is defined by the length of a single arm (see also the snapshots in Fig.~\ref{Re2_stars}).

\begin{figure}[h]
\centering
  \includegraphics[width=\columnwidth]{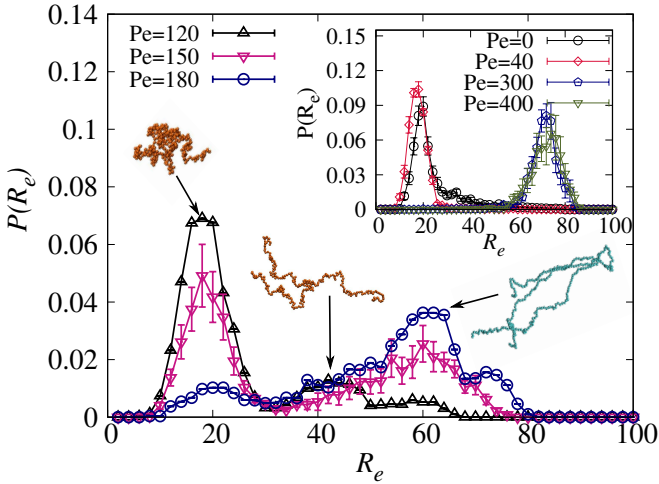}
  \caption{Probability distribution for the centre-to-end distance $R_e$ of a star polymer with $n_{a}=3$ and $N_{m}=100$ per arm.}
  \label{prob_star}
\end{figure}

To understand more about the stretching patterns shown by different active stars, it is essential to examine the local arrangements and the local density of beads as a function of activity. Fig.~\ref{mono_cnt} shows the dependence of the average number of nearest neighbours $\left<C_{n}\right>$ of a monomer on the corresponding architecture and $\mathrm{Pe}$. Nearest neighbours are counted within a radius of $3.5\sigma$ from a monomer. A gradual decline in the value of $\left<C_{n}\right>$ for small $\mathrm{Pe}$ values corresponds to slight stretching before the expansion transition, which is later followed by a sharp decrease in $\left<C_{n}\right>$ within the region of transition to the expanded state which starts at around $\mathrm{Pe}\approx 100$. The plot is consistent with our previous findings for $\frac{\left<R_{e}^{2}\right>_a}{\left<R_{e,0}^{2}\right>_a}$ explained above. One could also notice the increase in the error bar values for the extended conformations in comparison to the coil conformations (averaging was done for 20 different realisations each time). This shows that the shape and the set of conformations fluctuate more for high $\mathrm{Pe}$ numbers. If one normalises three sets of values shown in Fig.~\ref{mono_cnt} by the corresponding numbers for the passive case, $\left<C_{n}\right>/\left<C_{0}\right>$, then the data will merge into a single set within the accuracy of simulations (not shown here).

In order to clarify the conformational behaviour before, during and after the expansion transition we show the probability distribution function of centre-to-end distance $R_{e}$ of a star polymer with $n_{a}=3$ and $N_{m}=100$ per arm in Fig.~\ref{prob_star}. For the passive star polymers ($\mathrm{Pe}=0$) as well as for small $\mathrm{Pe}$ values  ($\mathrm{Pe}=40$) in the inset of Fig.~\ref{prob_star} we see a single peak in the distribution at small $R_e$, suggesting that all polymers are in a single compact form. For sufficiently higher values of $\mathrm{Pe}=300,400$ the peak shows a significant shift towards large values of $R_{e}$ (the inset of Fig.~\ref{prob_star}). This observation indicates two clearly distinctive conformations of active star polymer. The first one being a relatively compact coil while the second having a stretched shape. However, the most interesting case is observed at the intermediate values of active force strengths, $\mathrm{Pe}=120,150,180$ (see the main plots in Fig.~\ref{prob_star}), where we can see two coexisting conformations. This observation points out that the transition of star polymer from a compact coil geometry towards a stretched one passes through a region with two possible coexisting shapes.

To conclude the above observations, we can see that once the active force reaches a threshold value $(\mathrm{Pe}\approx200)$ in this model, the polymer arms are close to a maximum stretching through conformation change. A further stretching is not happening in the accessible range of parameters due to the fact that the connectivity of the beads is ensured by a relatively steep harmonic potential. This saturation is observed for stars as well as linear architectures and points out that the power-law characterises the expansion transition rather than scaling at large $\mathrm{Pe}$ numbers observed and discussed in the earlier literature. The nature of the transition is well shown by the probability distribution of a centre-to-end distance for star polymers (Fig.~\ref{prob_star}). At intermediate $\mathrm{Pe}$ numbers for star polymers, $\mathrm{Pe}=120,150,180$ in Fig.~\ref{prob_star}, they stay in a co-existing state just before achieving their complete stretched conformation.

\begin{figure}[h]
\centering
  \includegraphics[width=\columnwidth]{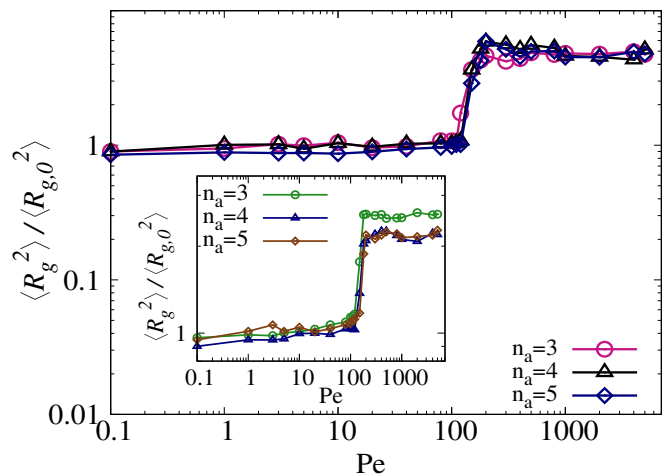}
  \caption{Scaled average squared radius of gyration $\frac{\left<R_{g}^{2}\right>}{\left<R_{g,0}^{2}\right>}$ as a function of $\mathrm{Pe}$ for active star polymers with number of arms $ n_{a}=3,4,5$, respectively. Main plot: $N_{m}=100$. Inset: $N_{m}=50$.}
  \label{Rg2_stars}
\end{figure}

\textbf{Expansion behaviour.} To determine how the active stars expand as a function of the applied active force, we calculate the average squared radius of gyration of stars $\left<R_{g}^{2}\right>$ with the number of arms $n_{a}=3,4,5$ as a function of $\mathrm{Pe}$  (Fig.~\ref{Rg2_stars}). Only a small deviation of $\frac{\left<R_{g}^{2}\right>}{\left<R_{g,0}^{2}\right>}$ from the value for the passive case is visible when $\mathrm{Pe}<100$. However, just as for the centre-to-end distance a significant growth after $\mathrm{Pe}\approx 100$ and finally a saturation starting for $\mathrm{Pe}\geq 200$ is visible for all the cases. The expansion for stars with the same molecular mass can also be investigated and the results are qualitatively consistent with the results for $\frac{\left<R_{e}^{2}\right>_a}{\left<R_{e,0}^{2}\right>_a}$. 

\begin{figure}[h]
\centering
  \includegraphics[width=\columnwidth]{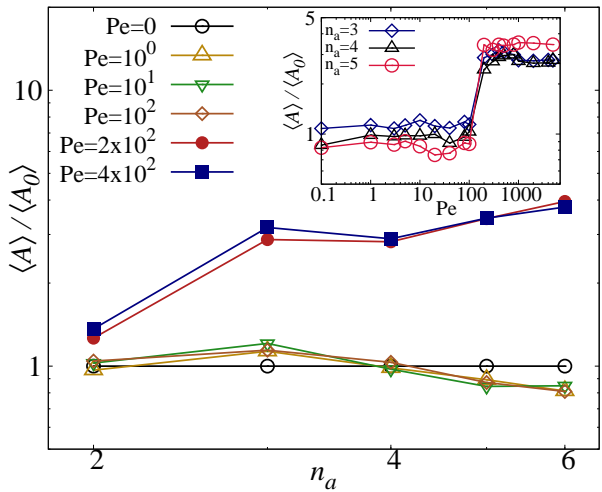}
  \caption{Asphericity of active star polymers with different arm numbers $n_{a}=3,4,5,6$ shown for various $\mathrm{Pe}$ in the main plot. The behaviour of each star as a function of $\mathrm{Pe}$ with $n_{a}=3,4,5$ is shown in the inset. Here, $N_{m}=100$.}
  \label{asp}
\end{figure}

\begin{table} [h!]
	\centering
	\begin{tabular}{ |p{1cm}|p{3.0cm}| p{3.0cm}| }
		\hline
		$\mathrm{Pe}$ & $n_{a}=3$ & $n_{a}=5$\\  
		\hline 
		0 & 0.069$\pm$0.0004 & 0.035$\pm$0.0004\\
		\hline
		1 & 0.078$\pm$0.0006 & 0.045$\pm$0.0004\\
		\hline
		40& 0.075$\pm$0.0005 & 0.038$\pm$0.0003\\
		\hline
		100& 0.079$\pm$0.0007 & 0.044$\pm$0.0005\\
		\hline
		200& 0.199$\pm$0.0005 & 0.173 $\pm$ 0.0007\\
		\hline
		500& 0.215$\pm$0.0002 & 0.209 $\pm$ 0.0003\\
		\hline
		1000&  0.170$\pm$0.0005 &  0.208 $\pm$ 0.0003\\
		\hline
	\end{tabular}
	\caption{Asphericity $\left<A\right>$ (not scaled) of stars with $n_{a}=3,5$ for various $\mathrm{Pe}$, $N_{m}=100$. }
	\label{table1}
\end{table}

Further, in Fig.~\ref{asp} we have plotted the scaled asphericity (w.r.t. the passive star with the same number of arms) of stars with different arm numbers $(n_{a}=3,4,5,6$ and $N_{m}=100)$ as a function of $\mathrm{Pe}$ (see also Table 1 for the values). For $\mathrm{Pe}\le 100$ the asphericity ratio $\left<A\right>/\left<A_0\right>< 1$ indicates that the active star conformations are closer to the spherical shape than for passive ones, while for $\mathrm{Pe}> 100$ the asphericity ratio $\left<A\right>/\left<A_0\right> > 1$ points out that the active stars are becoming more asymmetric. The nature of individual star's asphericity is displayed in the inset of Fig.~\ref{asp} for stars with $n_{a}=3,4,5$ and $N_{m}=100$. The plot indicates that the asphericity also has small variations in the low $\mathrm{Pe}$ regime, and a saturation followed by a significant increment for higher $\mathrm{Pe}$ values is observed. This suggests that the star polymers are initially in a relatively compact state, and in a higher force regime, they start swelling until the saturation is achieved. We summarise the asphericity in a table form for various active forces as in Table~\ref{table1} for star polymers with $n_{a}=3,5$. One can notice that the asphericity for $\mathrm{Pe}>100$ starts increasing for each case and achieves more or less a constant value for higher $\mathrm{Pe}$.

\begin{figure}[h]
\centering
  \includegraphics[width=\columnwidth]{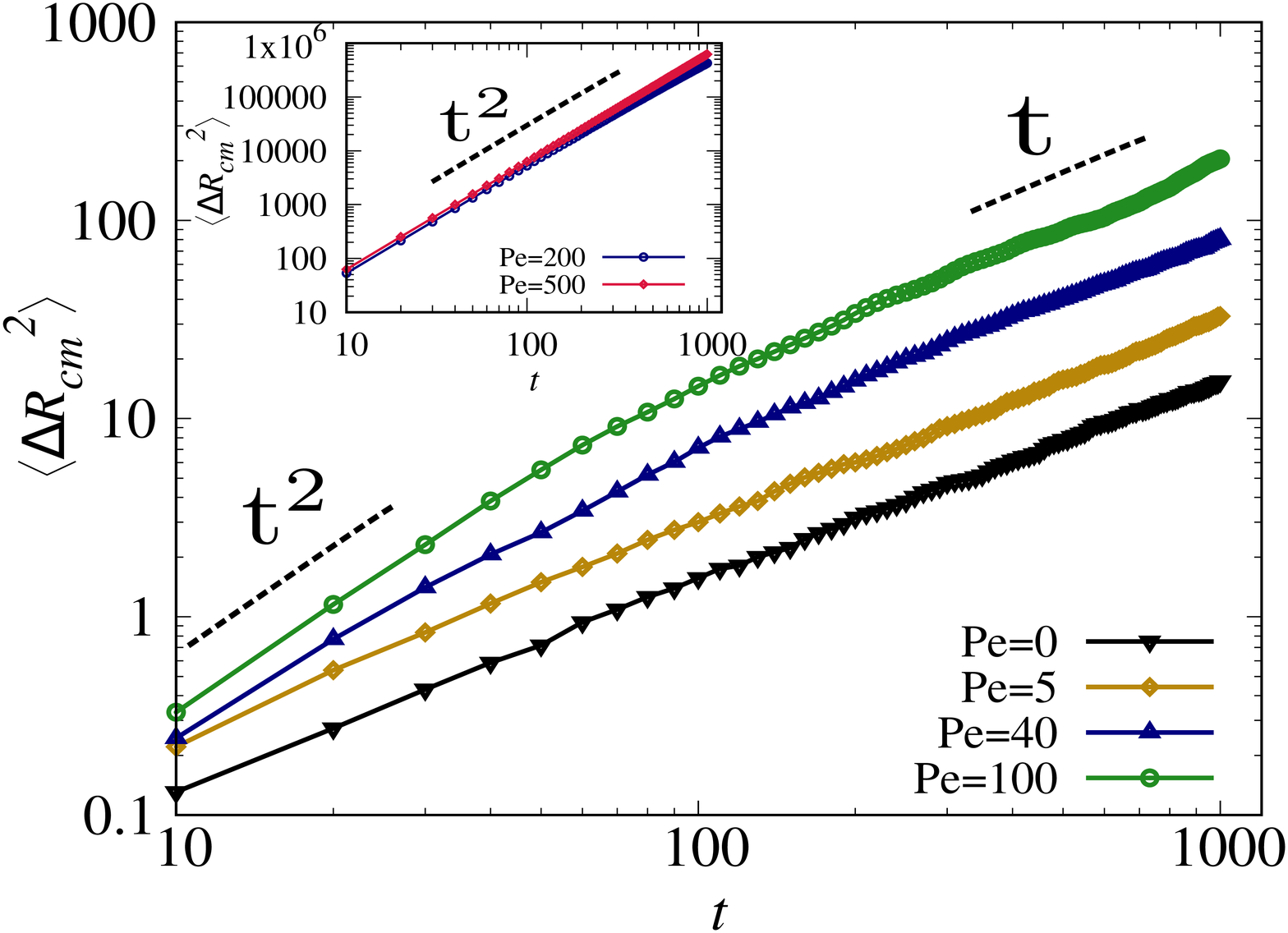}
  \caption{Mean-squared displacement of the centre of mass of an active star polymer with number of arms $n_{a}=3$ and number of monomers per arm $N_{m}=100$.}
  \label{msd_star}
\end{figure}

\textbf{Diffusion and dynamics}. The activity should affect the overall diffusion behaviour of active star polymers as a whole. Since these stars can transport the cargo and can be used for drug delivery~\cite{Sulistio:2011,Yang:2017, England:2020}, it is essential to analyse their dynamical behaviour. To see the diffusion pattern of active star polymers, we plot the mean-squared displacement $\left<\Delta {\bf{R}}_{\mathrm{cm}}^{2}\right>=\left<({\bf{R}}_{\mathrm{cm}}(t)-{\bf{R}}_{\mathrm{cm}}(0))^{2}\right>$ of the centre of mass of stars and compare it with the passive case. As we can see in Fig.~\ref{msd_star}, applying the activity to the polymer monomers enhances the diffusion. Therefore, the active star polymers have higher diffusivity than their passive counterparts. The plot shows that for $\mathrm{Pe}\le 100$ the scaling behaviour $\left<\Delta {\bf{R}}_{\mathrm{cm}}^{2}\right> \sim t^{2}$ reveals ballistic motion at short times while at long times the motion becomes Brownian $\left<\Delta {\bf{R}}_{\mathrm{cm}}^{2}\right> \sim t$. However, for $\mathrm{Pe}> 100$, we do not see the normal diffusion regime yet even for relatively higher times, which suggests that the active stars with high $\mathrm{Pe}$ show ballistic motion within the limit of our simulation time.

\section{Conclusions}\label{conclusion}

To summarise, we have assessed how branching of active polymer affects the behaviour under the increase of activity for the case of star polymers. For the active star polymers within the simulation model we observe a rather steep transition from coil to the extended conformation which does not allow us to solidly derive any scaling dependency for this transition. We reach the regime of higher $\mathrm{Pe}$ values than in previous studies and conclude that the previously reported scaling for the linear case is a feature of transition from the more compact coil state to the extended state rather than the law which characterises the unbounded growth of a polymer gyration radius which is a seeming conclusion from some of the earlier work. Moreover, by plotting the probability distributions of  end-to-end distances for the arms we see that during this transition two typical conformations could coexist including the coil-like conformation and the strongly stretched conformation. The work carried out here underlines the importance of architecture for active polymers and suggests a noticeable difference in behaviour. This motivates the community to investigate more about the branched polymer architectures and utilise the applications of such active structures in various fields. A profound example of such structures can be used in controlled drug release where one can take an advantage of the unique topological structures.

\begin{acknowledgments}
The authors acknowledge the use of computational resources of the Skoltech supercomputer Zhores for obtaining the results presented in this article.
\end{acknowledgments}


\section{Appendixes}

\appendix{}
\section{Linear active polymer simulations}
\begin{figure}[h]
\centering
  \includegraphics[width=\columnwidth]{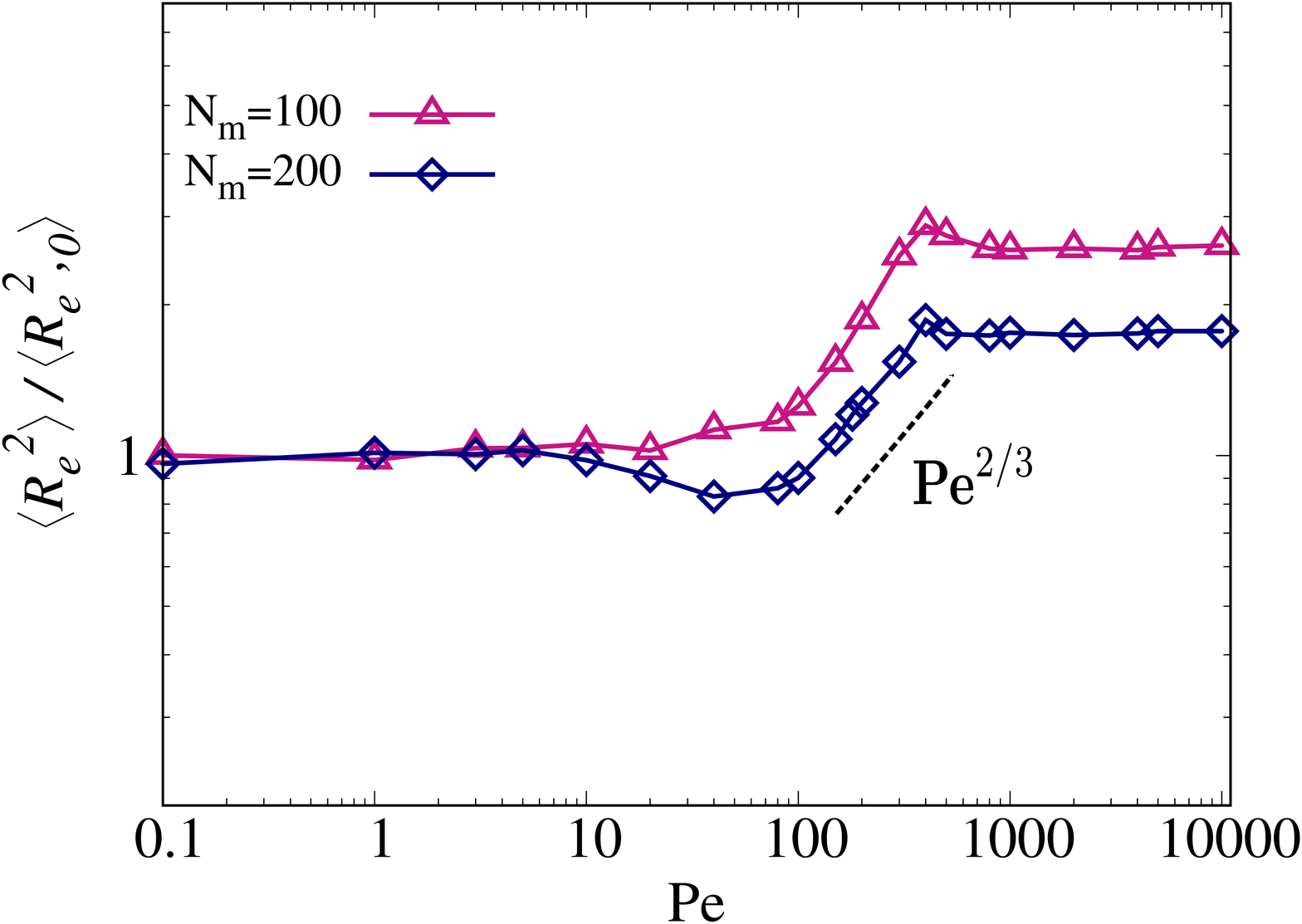}
  \caption{Scaled average squared end-to-end distance $\frac{\left<R_{e}^{2}\right>}{\left<R_{e,0}^{2}\right>}$ of linear active polymers as a function of activity, i.e. varying with $\mathrm{Pe}$.}
  \label{linear_ply}
\end{figure}

\begin{figure}[h]
\centering
  \includegraphics[width=\columnwidth]{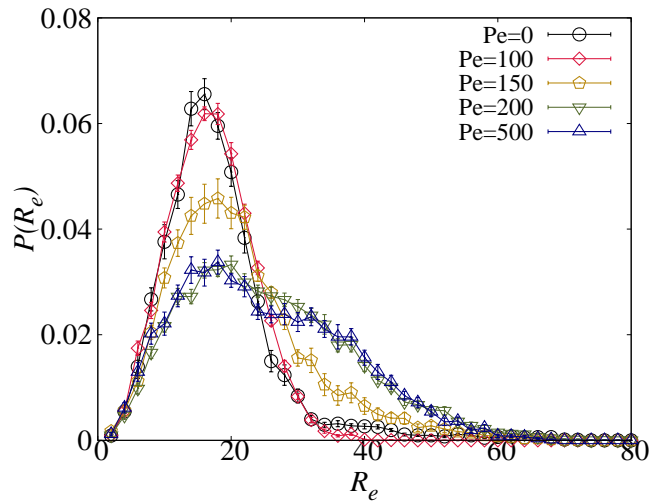}
  \caption{Probability distribution for the end-to-end distance $R_e$ of a linear polymer with $N_{m}=100$ as a function of activity, i.e. varying with $\mathrm{Pe}$.}
  \label{prob_lin}
\end{figure}

 The ratio of averaged square end-to-end distance at non-zero activity to the passive value or dimensionless stretching ratio for linear active polymer with numbers of monomers per arm $N_{m}=100,200$ is shown below in Fig.~\ref{linear_ply}. Although the transition range spreads just over half a decade only if one estimates the scaling exponent from that the earlier reported scaling $\mathrm{Pe}^{2/3}$ can be observed. However, contrary to many of the previously published results where the largest $\mathrm{Pe}$ values considered were about 200 we also simulate stronger active forces and observe a clear saturation plateau. This points out that the scaling characterises transition from the coil to the stretched state rather than actual behaviour at large $\mathrm{Pe}$ values.
 
In Fig.~\ref{prob_lin} we show the probability distribution functions for the linear chains for activities. One can clearly see that, contrary to Fig.~\ref{prob_star} the transition to the extended state proceeds continuously and without two separate coexisting confirmations (two separate peaks) within the expansion transition region.

\begin{figure}
    \centering
   \includegraphics[width=\columnwidth]{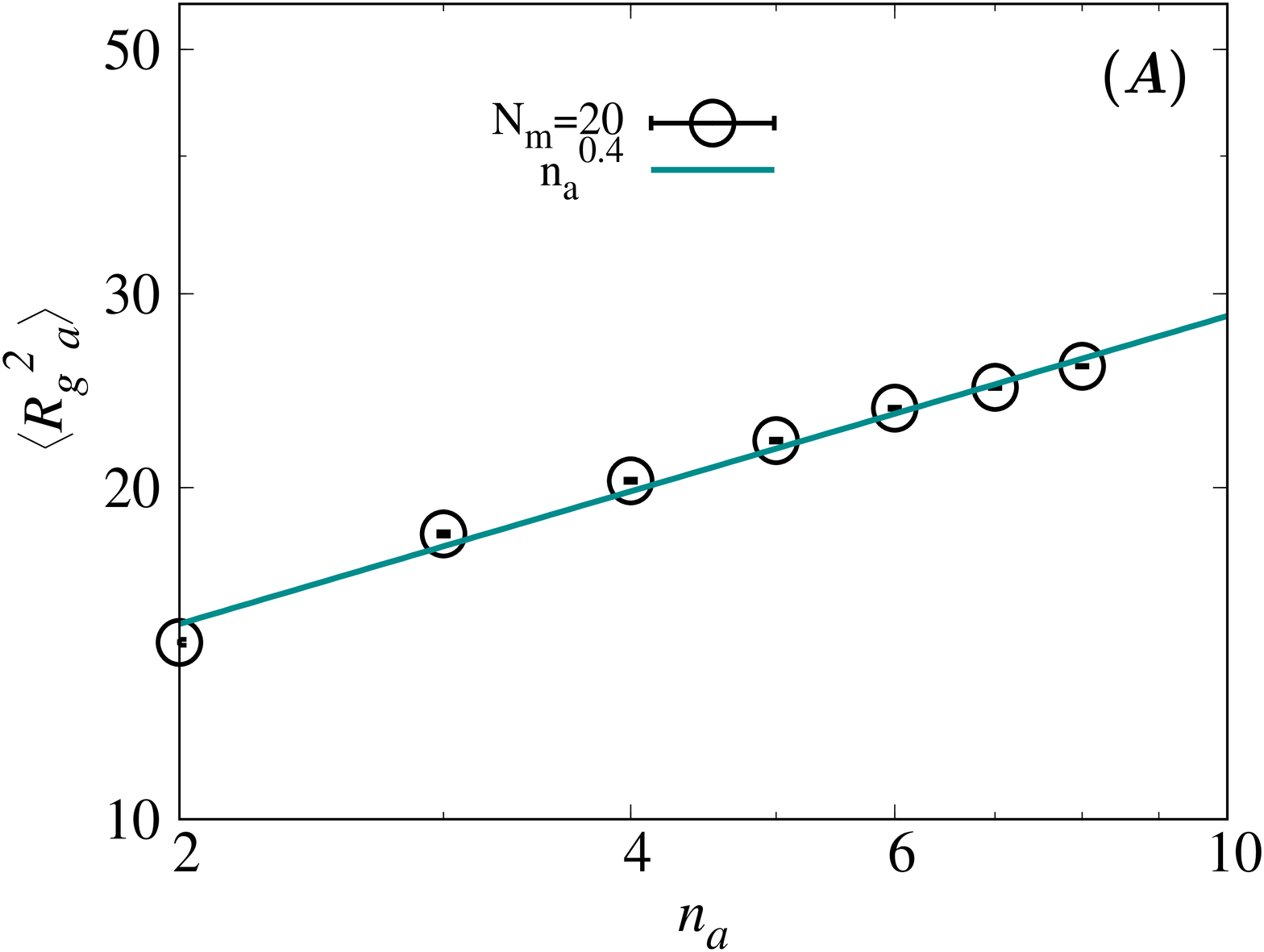}
   \includegraphics[width=\columnwidth]{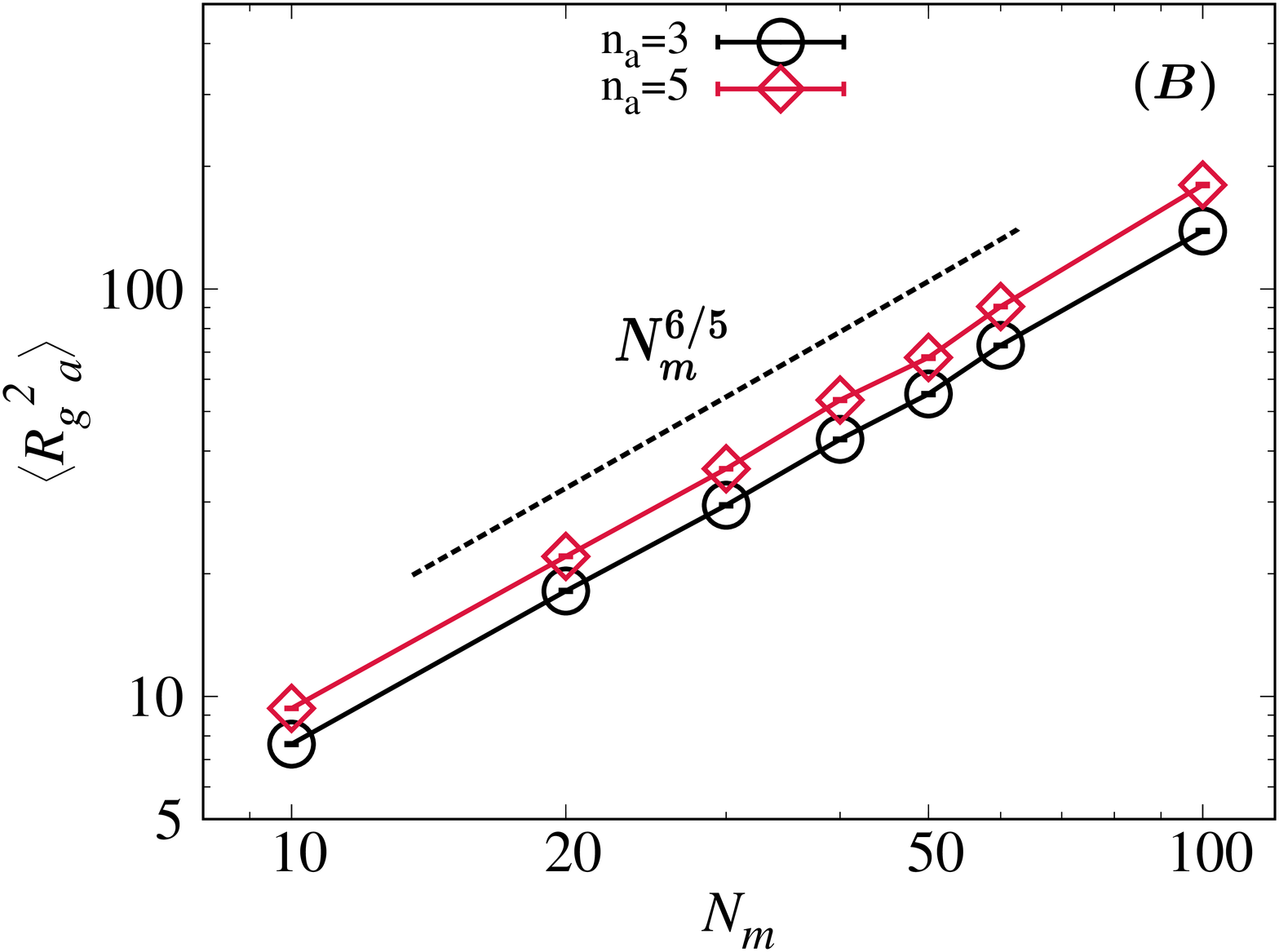}
    \caption{A. Average square radius of gyration for passive star polymers varying with arm number $n_a$, B. Average square radius of gyration for passive star polymers varying with arm length for passive stars with $n_{a}=3,5$.}
    \label{Rg2_passive}
\end{figure}

\section{Passive star polymer simulations}

In Fig.~\ref{Rg2_passive} we show that our protocol produces the known results for passive star polymers. Fig.~\ref{Rg2_passive}(A) presents the average squared radius of gyration $\langle{R_{g}^{2}}_{a}\rangle>$ varying with number of arms $n_{a}$ for a passive star having the number of monomer per arm as $N_{m}=20$. We reproduce the known scaling with the number of arms ${\langle R_{g}^{2}}_{a}\rangle \sim n_{a}^{1-\nu}$. Here, the exponent $\nu\sim 0.6$~\cite{Grest:1994,Kalyuzhnyi:2019,Khare:2014}. Also, in Fig.~\ref{Rg2_passive}(B), we show the dependence of average squared radius of gyration $\langle{R_{g}^{2}}_{a}\rangle$ of passive star polymers $n_{a}=3,5$ on their arm length $N_{m}$. The scaling follows the scaling for the polymer in good solvent, i.e. ${\langle R_{g}^{2}}_{a}\rangle \sim N_{m}^{2\nu}$ similar to a linear polymer chain~\cite{Duplantier:1989,Kalyuzhnyi:2019,Khare:2014}.

 \begin{figure}[h]
\centering
  \includegraphics[width=\columnwidth]{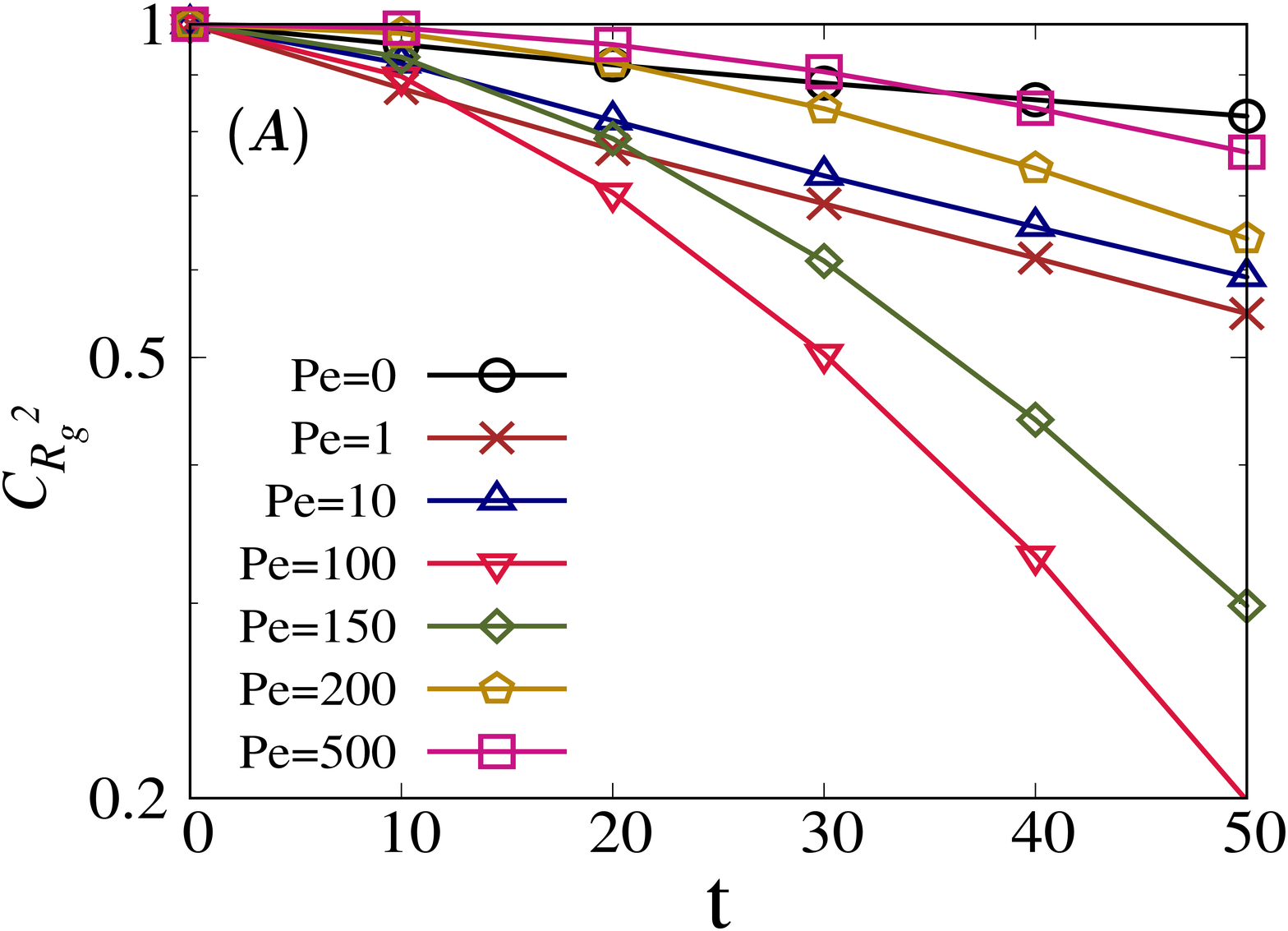}
  \includegraphics[width=\columnwidth]{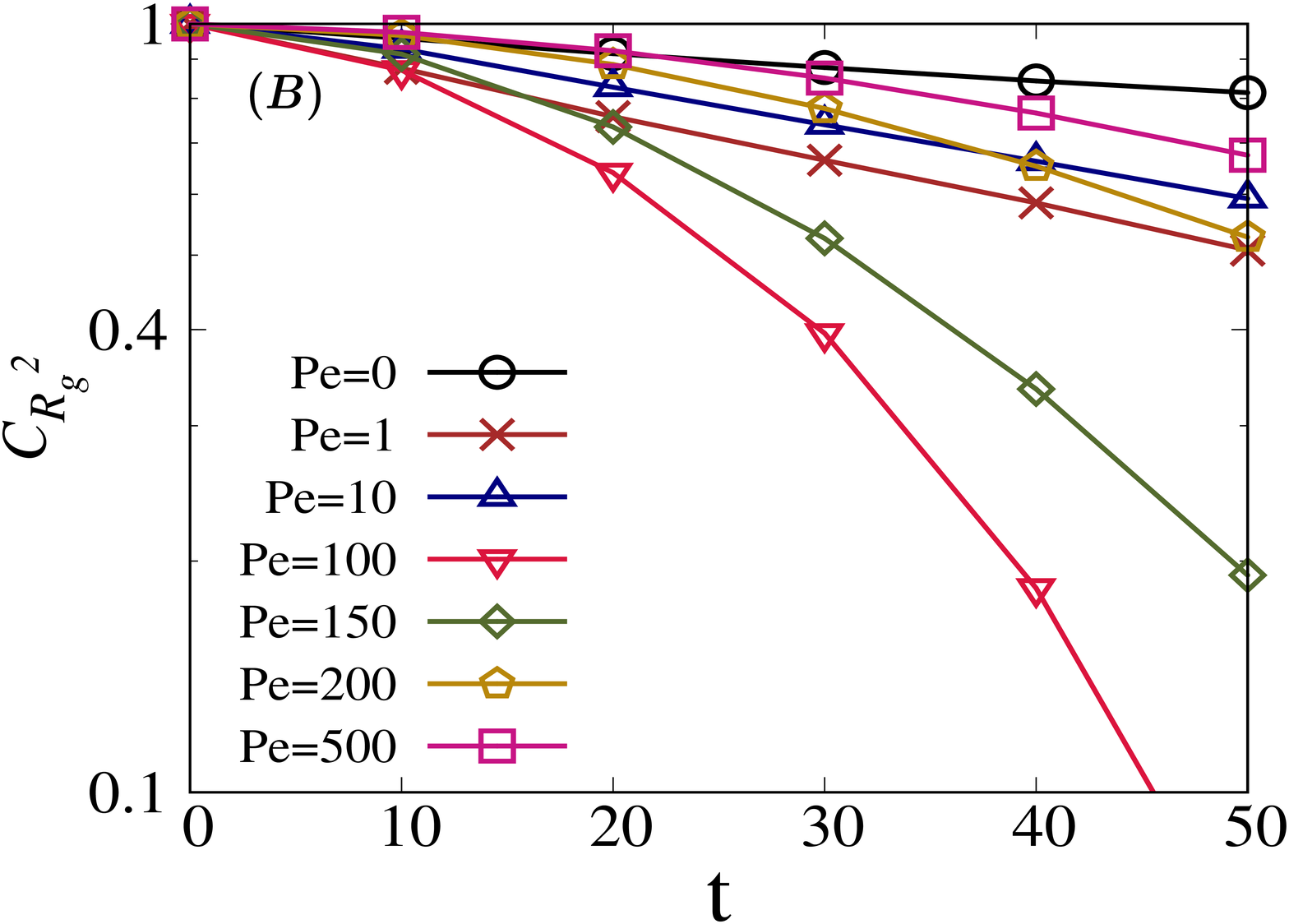}
  \caption{Time dependence of correlation function for the gyration radius $C_{{R}^{2}_g}$, A: For a linear chain with $N_{m}=100$, B: For a star polymer with $N_{m}=50$ and $n_{a}=3$.}
  \label{corr_linear_3star}
\end{figure}

\section{Correlation functions for the gyration radius}

If one wants to see the relaxation of conformations without including the orientational or positional relaxation of the polymer as a whole, one could study the correlation functions for the gyration radii according to the following equation~\cite{lyulin2004computer},
\begin{equation}
C_{R_g^2}(t)=\frac{\langle R_g^2(0)R_g^2(t)\rangle-\langle R_g^2\rangle^2}{\langle R_g^4\rangle-\langle R_g^2\rangle^2}.
\label{gyrationcorrelation}
\end{equation}
We have used two settings for the calculation of the correlation function: The relaxation from the fully stretched polymer and the relaxation from the fully collapsed, globular form. Qualitatively the results are the same in these cases. In Fig.~\ref{corr_linear_3star} we show the time dependence of correlation function for the squared radius of gyration $(C_{{R}^{2}_g})$ starting the from a globular form for linear and star polymers with 3 arms. We see that at low activities the correlation decays exponentially yet the relaxation time behaves non-monotonously in both cases. However, for higher activities, i.e. once we approach the expansion transition and beyond the multimode relaxation is seen and determination of a single relaxation time becomes difficult. Importantly, the same effect is observed in linear and star polymers for our simulation model and the qualitative behaviour of the gyration radius correlation for both of these two architectures is the same. We believe that this multimode relaxation at higher $\mathrm{Pe}$ numbers is a feature of the simulation model and shows its difference from the approaches such as in Ref.~\cite{Winkler:2017}. A similar complexity in terms of relaxation with the simulation model used by us was observed for ring polymers~\cite{Emanuele:2021}. From the Supplementary Information for the latter study it is clear that only for small $\mathrm{Pe}$ the relaxation happens in a single mode fashion.

\section{Squared centre-to-end distance of stars with the same molecular mass}

Here we add a plot which shows the average squared centre-to-end distance $\frac{\left<R_{e}^{2}\right>_a}{\left<R_{e,0}^{2}\right>_a}$  of star polymers when the molecular mass of polymers with different number of arms is kept constant. Naturally it means a decrease of the arm length with an increase of the arm number.

\begin{figure} [h]
    \centering
   \includegraphics[width=\columnwidth]{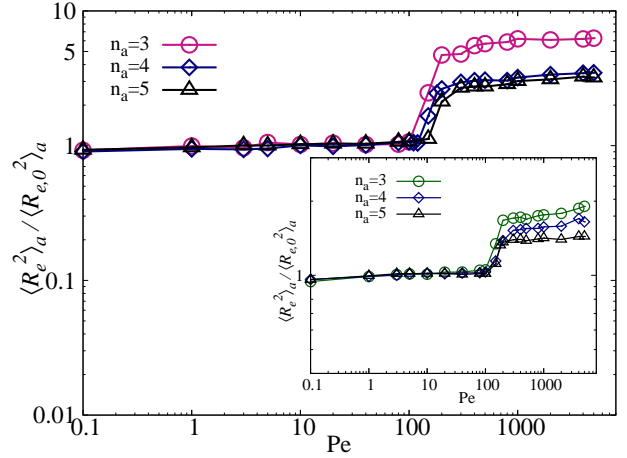}
    \caption{Average squared centre-to-end distance $\frac{\left<R_{e}^{2}\right>_a}{\left<R_{e,0}^{2}\right>_a}$  of star polymers with the same molecular mass, but different number of arms. The main plot is for longer polymer arms (total molecular mass $\approx 201m$) and the inset shows the stretching for shorter arms (total molecular mass $\approx 51m$).}
    \label{Re2_51_201}
\end{figure}


\nocite{*}

\bibliography{pre}

\end{document}